# MAGNETIC QUBITS AS HARDWARE FOR QUANTUM COMPUTERS.

J. Tejada*, E. M. Chudnovsky†, E. del Barco*, J. M. Hernandez* and T. P. Spiller‡

*Physics Department. University of Barcelona, Diagonal 647, 08028 Barcelona, Spain.*

*† Lehman College of the City University of New York, Bedford Park Boulevard West, NY 10468-1589.*

*‡ Hewlett-Packard Laboratories, Filton Road, Stoke Gifford, Bristol BS34 8QZ, UK.*

**We propose two potential realisations for quantum bits based on nanometre scale magnetic particles of large spin S and high anisotropy molecular clusters. In case (1) the bit-value basis states |0⟩ and |1⟩ are the ground and first excited spin states $S_z = S$ and $S-1$, separated by an energy gap given by the ferromagnetic resonance (FMR) frequency. In case (2), when there is significant tunnelling through the anisotropy barrier, the qubit states correspond to the symmetric, |0⟩, and antisymmetric, |1⟩, combinations of the two-fold degenerate ground state $S_z = \pm S$. In each case the temperature of operation must be low compared to the energy gap, $\Delta$, between the states |0⟩ and |1⟩. The gap $\Delta$ in case (2) can be controlled with an external magnetic field perpendicular to the easy axis of the molecular cluster. The states of different molecular clusters and magnetic particles may be entangled by connecting them by superconducting lines with Josephson switches, leading to the potential for quantum computing hardware.**

It is known already from the existence of quantum algorithms, such as for factoring[1] or searching[2], that technology which stores and processes information according to the laws of quantum physics will be capable of computational tasks infeasible with any conventional information technology. Almost all current work on quantum computing has focussed on quantum bits, two-state quantum systems, or qubits (although in principle larger individual Hilbert spaces could be used). A useful factoring or searching quantum computer will require many qubits (e.g. about $10^5$ for factoring with error correction), although other applications may be realised with fewer. Quantum information processing in general (for background, see e.g. refs. 3–8) is currently a rapidly developing interdisciplinary field, embracing theory and experiment, hardware and software. This paper discusses a new magnetic approach for quantum hardware.

Five criteria which must be satisfied by candidate quantum computing hardware have been elucidated[9]: (i) Clearly identifiable qubits (an enumerable Hilbert space) and the ability to scale up in number; (ii) "Cold" starting states (e.g. the ability to prepare the thermal ground state of whole system); (iii) Low decoherence (so that error correction techniques[10,11] may be used in a fault-tolerant manner[12,13]) — an approximate benchmark is a fidelity loss of $10^{-4}$ per elementary quantum gate operation; (iv) Quantum gates (the ability to realise a universal set of gates through control of the system Hamiltonian); (v) Measurement (the ability to perform quantum measurements on the qubits to obtain the result of the computation). Any candidates for quantum computing hardware should be assessed against this "DiVincenzo checklist".

Over the last few years, a number of two-level systems have been examined (theoretically and/or experimentally) as candidates for qubits and quantum computing. These include ions in an electromagnetic trap[14–16], atoms in beams interacting with cavities at optical[17] or microwave[18] frequencies, electronic[19] and spin[20] states in quantum dots, nuclear spins in a molecule in solution[21,22] or in solid state[23], charge (single Cooper pair) states of nanometre-scale superconductors[24,25], flux states of superconducting circuits[26–28], quantum Hall systems[29] and states of electrons on superfluid helium[30]. All these systems score well on some aspects of the checklist; however, some open questions remain. For example, in general, decoherence looks to be more of a problem for the systems with the most promise of scalability. One thing is certain; there is currently no clear quantum computing favourite, analogous to the use of photons (down optical fibres or possibly free space) for quantum cryptography and communication. Consequently, in addition to further work on existing systems, new candidates for quantum computing hardware should be explored. This paper discusses possible magnetic cluster or particle realisations for qubits.

**Mesoscopic magnets and the spin Hamiltonian**

Magnetic particles of nanometre dimensions (both antiferromagnetic and ferrimagnetic) have two necessary ingredients to be considered as good candidates for quantum bits. One is a relatively large total spin S (a few hundred in units of Planck's constant in the case of very small antiferromagnetic particles); such spins are easier to prepare and measure compared to fundamental spins. The second is their high magnetic anisotropy; the two lowest levels inside the potential well of the anisotropy barrier, as per the case of the magnetic clusters, may be separated by an energy gap of a few kelvin (see Figure 1). Quantum dynamics of the magnetisation in antiferromagnetic particles was first theoretically suggested[31] and has also been established in experiment[32-34]. The existence of a low temperature resonance in the absorption spectrum of horse spleen ferritin has been attributed to quantum coherence[32,33].

Magnetic clusters such as $Fe_8$ are systems that can be described by metastable wells. The value of the spin, S=10, in these molecular clusters allows examination of the border between quantum and classical mechanics[35-37]. A molecular cluster has all the attributes of a mesoscopic system; it consists of thousands electrons and nucleons. The parameters (most importantly the energies) of the magnetic states of clusters are defined by the collective motions of all their constituent particles. Control of these parameters can be effected easily by external magnetic fields.

To a first approximation the spin Hamiltonian of both nanometre scale magnetic particles and magnetic clusters is:

$$H = -D S_z^2 + H' + H_{dis} \qquad (1)$$

Here D is the anisotropy constant, S is the spin of the particle/molecule, H' is the part of the Hamiltonian that introduces tunneling, $H_{dis}$ represents the interaction of the spin system with other magnetic units and environment degrees of freedom, z refers to the easy axis direction. The first term of the above Hamiltonian generates spin levels $S_z$ inside each well, separated by an energy $D(2S_z-1)$ which may be about 10 K in temperature units. In zero magnetic field, the spin levels in the two wells (which are separated by the barrier height $U=DS^2$) are degenerate. The longitudinal component of an applied magnetic field tilts the potential selecting the spin level in one of the two wells. The transverse component of the magnetic anisotropy can induce rapid resonant tunnelling transitions[37-45] between the degenerate spin levels of the two wells (see Figure 2), which can lead to superposition of the two spin levels[46]. Quantum tunnelling through the barrier leads to energy eigenstates that are superposition of spin states on either side of the barrier. The energy gap, the so-called tunneling splitting Δ, between two such levels is always much lower than the energy difference between the spin levels $S_z$. Weakly non-compensated antiferromagnetic particles can exhibit a significant tunneling splitting[37]. This tunnelling splitting may be of the order of several hundred millikelvin.

We propose that such magnetic systems have potential use for quantum computing hardware. A schematic example of the realisation is given in Figure 3. The magnetic qubits (clusters/particles) are arranged in a 1D lattice and coupled to the superconducting loops of micro-SQUID (Superconducting Quantum Interference Device) circuits as shown. In principal this arrangement could be extended to a 2D lattice. We discuss the relevant aspects of this system in more detail by addressing the five criteria in the hardware checklist[9].

**DiVincenzo checklist**

Identifiable qubits: In a fabricated magnetic system of many particles/clusters, every molecular cluster/particle can act as an individual qubit, identifiable through their spatial location. There are two potential realisations of a qubit. Case (1), detailed in the inset of Figure 1, corresponds to the situation when the quantum tunnelling of the spin of the molecular cluster/particle is suppressed. In this case the classical picture of the $|1\rangle$ state (the excited state) is the uniform precession of the magnetic moment of the particle about the direction of the effective field that is formed by the magnetic anisotropy of the particle and the external magnetic field. This excited state is separated from the ground state, $|0\rangle$, by an energy gap set by the FMR frequency, which can be around 1~10 K. For the example with total spin S = 10, the qubit state $|0\rangle$ is the ground state with $S_z = 10$ and the state $|1\rangle$ that with $S_z = 9$. Case (2), detailed in the inset of Figure 2, corresponds to the situation when there is significant spin tunnelling through the anisotropy barrier. The qubit states then correspond to the symmetric, $|0\rangle$, and antisymmetric, $|1\rangle$, combinations of the two-fold degenerate ground state $S_z = S$. The energy splitting between the qubit states depends on the spin tunnelling frequency and can be tuned using an external magnetic field perpendicular to the easy axis of the molecular cluster[46,47].

In the experiments already performed to investigate quantum behaviour in magnetic systems[37-45], non-interacting ensembles were used to obtain detectable signals. This is clearly inadequate for magnetic quantum computation with one qubit per cluster (unlike in the case of liquid state NMR[21,22], where the whole, albeit small, computer resides in each member of the molecular ensemble). For a qubit per cluster, the clusters must be arranged on a substrate in a controlled manner. (Such control over the cluster layout is also needed for gates between qubits and measurement of individual qubits.)

In this particular case, each nanometre-size particle or molecular cluster is first deposited in a well controlled position on a dielectric substrate, for example embedded within a large solid matrix, and inside a micro-SQUID (micro superconducting quantum interference device) loop. The quantum states of each qubit are manipulated and measured by sending and receiving electromagnetic signals to and from the corresponding micro-SQUID.

State preparation: State preparation (and indeed the operation of magnetic cluster qubits) must be realised at a temperature significantly lower than the energy gap, $\Delta$, between the states $|0\rangle$ and $|1\rangle$. The typical gap for case (1) is of the order of a few kelvin, while in case (2) the gap may be controlled by applying an external magnetic field perpendicular to the anisotropy axis of the molecular cluster. The operation of a magnetic cluster qubit must therefore be done in the kelvin down to millikelvin temperature range. A system of case (1) qubits allowed to attain equilibrium at such a temperature is thus effectively prepared in the state $|0000...\rangle$. Clearly the actual system state will be a thermal density operator, so there will be small contributions with excited qubits (and much smaller corrections with the cluster spin(s) in states outside the truncated space which forms the effective qubit(s) Hilbert space). These corrections can be calculated from the system parameters and the temperature and will be damped by an appropriate Boltzmann factor.

A case (1) qubit in state $|0\rangle$ subject to a sudden change in external magnetic field perpendicular to the easy axis of the molecular cluster (which changes the tunnelling magnitude to that appropriate for case (2)) is effectively subject to a sudden basis change.

In the new basis the state is an equal weight superposition of |0⟩ and |1⟩. Such superposition preparation has already effectively been achieved for non-interacting ensembles[43,44].

Decoherence: The state of any quantum computer system must not decohere too rapidly away from its desired (unitary) computation. The benchmark at the building block level is that the loss in fidelity has to be small during an elementary one- or two-qubit gate. For the former, an estimate of this loss can be made from the reciprocal of the quality factor (Q) for coherent quantum oscillations. The decoherence phenomena which are mostly responsible for the damping of spin quantum coherence in magnetic clusters are the couplings of S to the crystalline lattice, nuclei and electromagnetic fields. The resonance experiments[46,47] which have been performed at different temperatures and frequencies suggest that the spin quantum coherence can be maintained for at least $10^{-8}$ s in non-interacting ensembles of magnetic qubits. This indicates a Q of about $10^4$, which compares well with those for other quantum hardware and the error correction/fault-tolerance benchmarks. The quality factor of the FMR in pure dielectric ferromagnetic crystals can be as high as one million, which compares even more favourably with these benchmarks. The FMR and its quality factor have not yet been measured in individual nanoparticles. However, there is no reason to believe that these should be lower than in large crystals, provided that the interaction of the magnetic moment of the molecular cluster with the substrate on which it is deposited introduces only comparable decoherence to that in the crystal environment.

Selection rules for spins due to time reversal symmetry should be taken into account when considering the decay of the excited state |1⟩ into the ground state |0⟩ as these two states have different symmetry with respect to time reversal. That is, only time-odd spin Hamiltonian terms contained in $H_{dis}$ will contribute to the decoherence. The spin-phonon interaction which comes from the spin-orbit coupling is a time-even operator, therefore, the decay of the excited state $|1\rangle$ into $|0\rangle$ cannot occur through the spontaneous emission of a phonon.

Nuclear spins always destroy the coherence at zero field and must be eliminated from magnetic qubits by isotopic purification. Similarly, the presence of free non-superconducting electrons in the sample will decohere tunneling through the spin scattering of electrons, $D_e \propto s \cdot S$. Although this operator is time-even, free electrons incidentally passing through the magnetic particle/cluster, perturb $|0\rangle$ and $|1\rangle$, breaking their properties with respect to time reversal. Thus, strongly insulating materials should be chosen for magnetic qubits. The effect of incidental phonons due to, e.g., relaxation of elastic stress in the matrix (the 1/f noise), should be similar to the effect of incidental electrons in perturbing the wave function. Thus, the perfection of the lattice should be given serious thought when manufacturing magnetic or any other qubits. One should then worry about the decohering effect of spin interactions that are odd with respect to time reversal. These are Zeeman terms due to magnetic fields. For example, $H_{dis} = -g\mu S_z H(t)$ has a non-zero matrix element between $|0\rangle$ and $|1\rangle$. The effort should be made, therefore, to shield the magnetic qubit from unwanted magnetic fields during the process of quantum computation. This can be done by placing the magnetic particle/cluster inside a superconducting ring.

Very recently[48] the decoherence mechanisms in crystals of weakly anisotropic magnetic molecules has been discussed. The promising conclusion is that the quantum coherence of such molecules (e.g. $V_{15}$) is not suppressed below temperatures of the order of 1 K.

The decoherence introduced through the external source terms shown explicitly in the Hamiltonian (1) (i.e. due to the fluctuations in the magnetic fields applied to individual qubits) and through the microSQUID measurement environment must be kept small. The effective quality factor of individual magnetic qubits must be of the same order as those already observed for ensembles.

Quantum gates: In order to perform arbitrary quantum computations, it must be possible to realise a universal set of gates. It is known[49,50] that arbitrary single qubit operations and a two-qubit gate capable of generating maximal entanglement from a product state form a universal set. Such gates can be performed with magnetic qubits.

The realisation of single magnetic qubit gates can be performed by effectively producing Rabi oscillations between the ground and the first excited state. The frequency range, time interval between pulses and control are now feasible with current technology. In more detail, magnetic qubits can be manipulated in an analogous manner to the manipulation of spins in ensemble NMR quantum computation. Application of a resonant pulse effects a rotation (proportional to the time-field product) about an axis in the x-y plane set by the phase. (Rotations about the z-axis can be achieved through a combination of x- and y- rotations.)

The realisation of two-qubit gates can be realised by coupling the neighbouring magnetic clusters/particles through inductive superconducting loops, illustrated in Figure 3. This is an effective scalar interaction and dominates over the direct dipole-dipole interaction (which falls off with the cube of the separation distance). The interaction Hamiltonian for the loop arrangement of Figure 3 is $H_{int} = JS_{z,a}S_{z,b}$. The coupling J depends on the supercurrent induced in the loop by one spin and the field this produces at the site of the other. The advantages of using superconducting loops are: (i) Low dissipation; (ii) The ability to control the coupling. Josephson switches can be used to turn on the inductive loop coupling between qubits when it is required. Josephson junctions have already been applied as building blocks for classical digital circuits, since they can switch in extremely short times. Evolution under the action of the $H_{int}$ constitutes a coupled two-qubit rotation. This can effect a controlled phase shift and, when sandwiched between suitable single qubit rotations, thus be used to generate a CNOT gate.

The realisation of arbitrary single-qubit gates and CNOT constitute a universal set, so magnetic systems are suitable for the implementation of general quantum algorithms and error correction procedures.

Measurement: The measurement of magnetic qubits can be effected by their coupling to individual microSQUIDs, as illustrated in Figure 2. One operation mode of a SQUID device is as a very sensitive magnetometer, so with pick-up loops as shown a microSQUID device can effect a measurement of a qubit spin, projecting onto the $S_z$ basis. Current state of the art with microSQUID technology permits the measurement of spins down to the scale of S=10000. This sensitivity needs to be reduced by a further two or three orders of magnitude to enable the detection of individual magnetic cluster/particle qubits. Nevertheless, this is not an impossible goal for future microSQUID technology.

In this paper we have proposed magnetic realisations for quantum bits. Through discussion of the "DiVincenzo checklist" for quantum hardware, we have proposed how these qubits may be prepared, evolve and interact under external control to realise a universal set of gates and be measured. Such fabricated magnetic cluster/particle systems are therefore suitable candidates for quantum computing hardware. The fabrication of magnetic qubit arrays offers the promise of scaling up in qubit number, in a similar manner to other solid state proposals.

There is clearly similarity in concept between the proposed magnetic cluster/particle qubits (and their manipulation) and the effective spin qubits employed in NMR quantum computation. In the NMR case, the qubits are manipulated through the application of appropriate radio-frequency fields. This is extremely difficult to implement at the level of individual spins and so experiments to date have been restricted to bulk ensembles, although Kane's proposal[23] suggests a future route for addressing individual qubits. Existing experimental investigations on magnetic systems have certainly illustrated some of the necessary quantum coherence properties; however, these have also been for ensembles. Despite this, we believe that the probing and manipulation of individual magnetic cluster/particle qubits is within reach using microSQUIDs because typical FMR frequencies and typical microSQUID frequencies fall in the same microwave range. As with other solid state candidates for quantum computing hardware, experiments are currently behind those on bulk NMR and ion trap systems. Nevertheless, we believe that magnetic cluster/particle systems are good candidates for further experimental and theoretical investigation.

E. M. C. acknowledges support by the National Science Foundation.

**Correspondence and requests for material should be addressed to J. T. (e-mail: jtejada@ubxlab.com).**

Figure 1. Case (1) qubit (detailed in the inset). The quantum tunnelling of the spin of the molecular cluster/particle is suppressed. $|0\rangle$ is the ground state in one of the wells and $|1\rangle$ is the first excited state, separated by an energy gap set by the ferromagnetic resonance (FMR) frequency. For the example with total spin S = 10, the qubit state |0⟩ is the ground state with $S_z$ = 10 and the state |1⟩ that with $S_z$= 9.

Figure 2. Case (2) qubit (detailed in the inset). There is significant spin tunnelling through the anisotropy barrier. The qubit states are the symmetric, |0⟩, and antisymmetric, |1⟩, combinations of the two-fold degenerate ground state $S_z$ = S. The energy splitting between the qubit states depends on the spin tunnelling frequency and can be tuned using an external magnetic field perpendicular to the easy axis of the molecular cluster.

Figure 3. A schematic example of the coupled controlled qubit realisation. The magnetic qubits (clusters/particles, yellow) are arranged in a 1D lattice and coupled to the superconducting loops of micro-SQUIDs (Superconducting Quantum Interference Devices, red) circuits as shown. The coupling circuits (blue) contain Josephson switches (green).

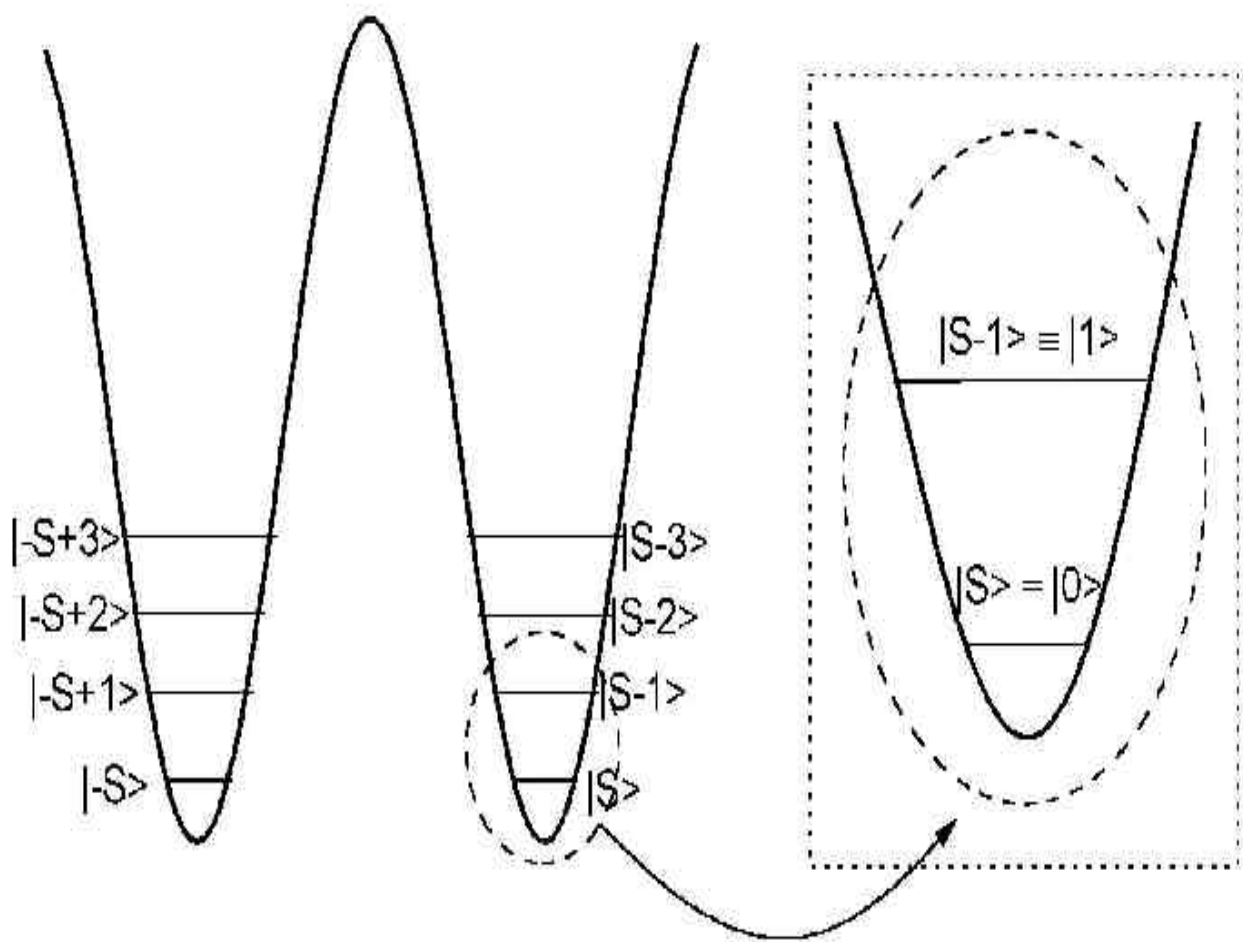

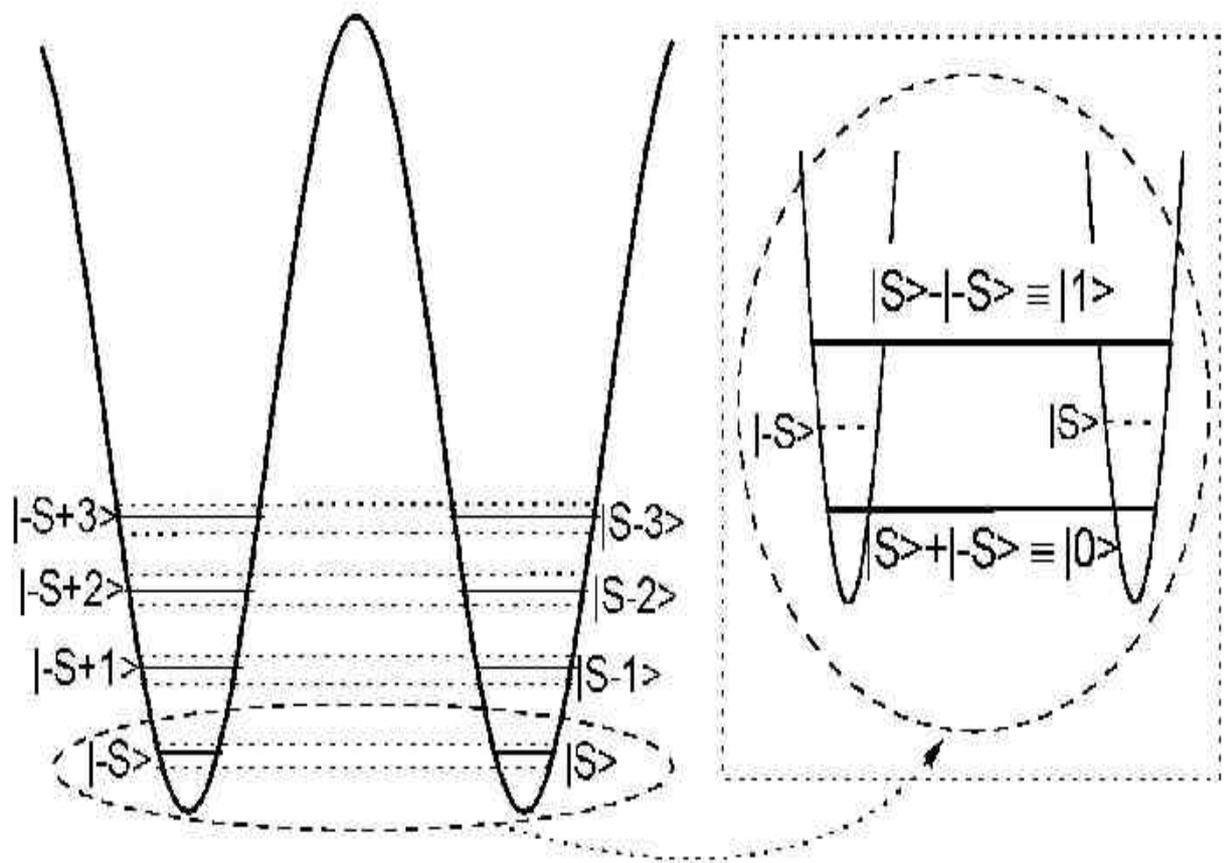

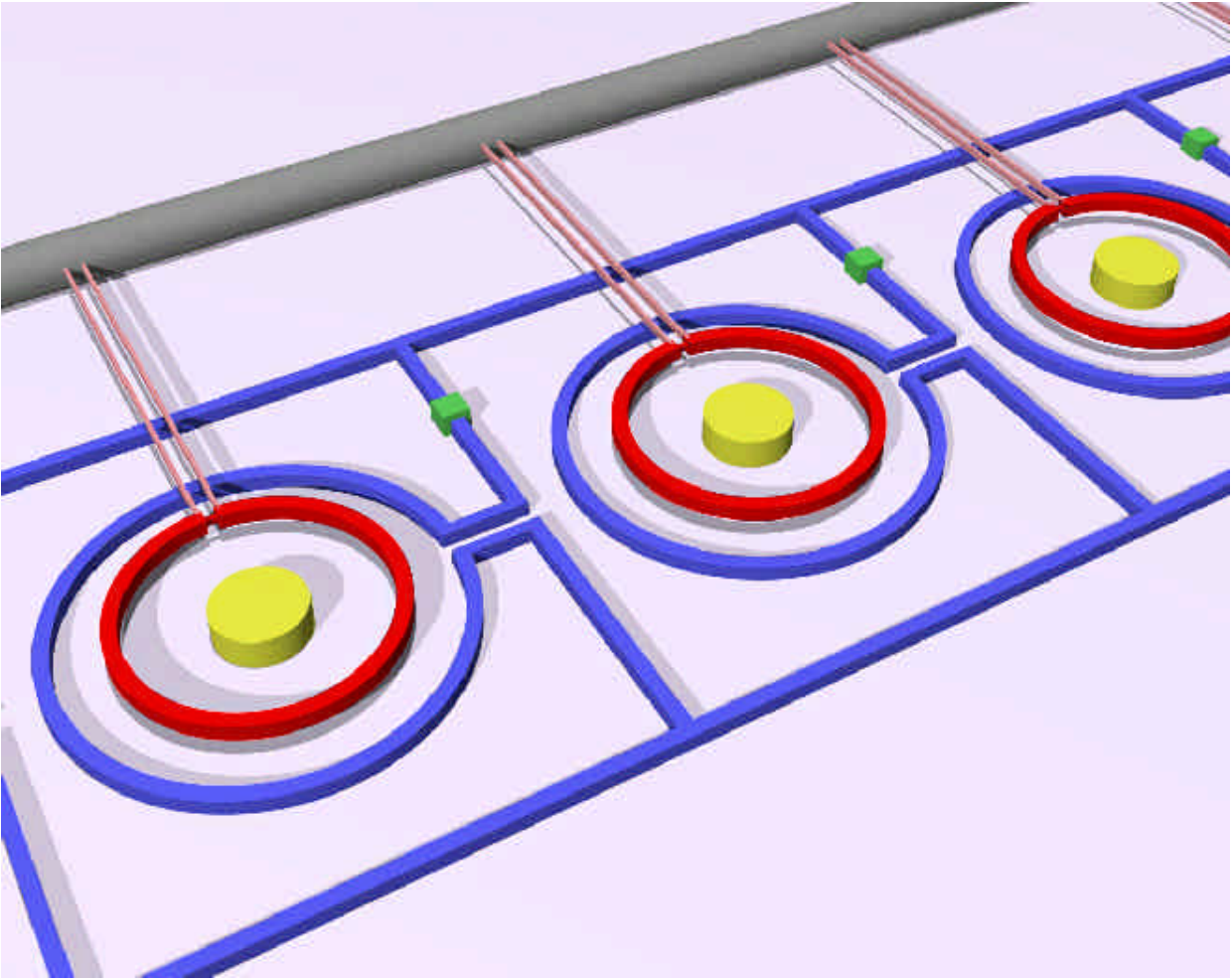